\documentclass[sigconf]{acmart}
\renewcommand\footnotetextcopyrightpermission[1]{} 
\usepackage{pifont} 

\usepackage{hyperref}
\usepackage{float}
\usepackage{tabularx}

\def\BibTeX{{\rm B\kern-.05em{\sc i\kern-.025em b}\kern-.08emT\kern-.1667em\lower.7ex\hbox{E}\kern-.125emX}}
    
\begin{document}

%
\title{Architectural Tactics to Improve the Environmental Sustainability of Microservices: A Rapid Review}

%


\author{Xingwen Xiao}
\affiliation{%
  \institution{University of Amsterdam, The Netherlands}
}
\email{xingwen.xiao@student.uva.nl}
\affiliation{%
  \institution{Vrije Universiteit Amsterdam, The Netherlands}
}
\email{x3.xiao@student.vu.nl}

%

%
\begin{abstract}

Microservices are a popular architectural style adopted by the industry when it comes to deploying software that requires scalability, maintainability, and agile development. There is an increasing demand for improving the sustainability of microservice systems in the industry. This rapid review gathers 22 peer-reviewed studies and synthesizes architectural tactics that improve the environmental sustainability of microservices from them. We list 6 tactics that are presented in an actionable way and categorized according to their sustainability aspects and context. The sustainability aspects include energy efficiency, carbon efficiency, and resource efficiency, among which resource efficiency is the most researched one while energy efficiency and carbon efficiency are still in the early stage of study. The context categorization, including serverless platforms, decentralized networks, etc., helps to identify the tactics that we can use in a specific setting. Additionally, we present how the evidence of optimization after adopting these tactics is presented, like the measurement unit and statistical methods, and how experiments are generally set up so that this review is both instructive for our future study and our industrial practitioners' interest. We further study the insufficiencies of the current study and hope to provide insight for other researchers and the industry.

\end{abstract}

%
%


%

%

%
\maketitle

\section{Introduction}
\label{sec-intro}

Microservices, in contrast to the traditional monolithic approach for deploying cloud services, break down the service into smaller, flexible chunks that can be combined. The decomposition of services can vary depending on the context. Some microservices function as API endpoints, while others serve as intermediaries for communication between different services \cite{msecommerce}.

Large technology firms are implementing this type of architectural design due to numerous benefits, including scalability and maintainability \cite{di2018migrating}. According to a survey conducted in 2020 \cite{OReillyRadar2020}, over half of the professionals from various roles within the industry have been utilizing microservices for more than a year. Moreover, there is a growing trend of learning and deploying microservices within the sector \cite{IBMCloud2021}. However, with the increasing popularity of microservices, there are significant non-negligible challenges. One such emerging concern is energy efficiency.

The ICT industry's sustainability issue has gained more attention in recent years due to its significant contribution to greenhouse gas emissions, accounting for 2.3\% of global carbon dioxide emissions in 2020. Additionally, the industry faces infrastructure costs as it continues to develop \cite{radu2017green}. Various countries have implemented green computing laws, such as the Ecodesign Directive and Directives on Ecolabeling in the European Union \cite{lohse2015law}. The enforcement of these laws remains a concern for the industry. Given the ongoing climate change, it is likely that more legislation regarding IT sustainability will be introduced in the future.

Due to the growing popularity of microservices in the ICT industry and the pressures companies face from climate, economy, and legislation, there is a need to investigate ways to improve the environmental sustainability of microservices.


Our objective is to identify architectural tactics that enhance the environmental sustainability of microservices. Architectural tactics are the architectural design decisions that are made to adjust a quality attribute model by changing some of its underlying variables \cite{bachmann2003deriving}. We will explain this in detail in Section \ref{sec:related_work}.

Following the rapid review approach \cite{cartaxo2020rapid}, we address research questions regarding the effectiveness of new approaches. We will explain the difference between a rapid review and a regular systematic literature review, as well as the motivation behind our choice in Section \ref{sec:research-goal}. To expedite the search strategy, we focus on a limited number of sources. We then provide concise and comprehensible summaries of the tactics mentioned in the selected research for future studies and practical application.

In this review, we explore research to synthesize architectural tactics specifically designed for optimizing microservices in terms of environmental sustainability. We propose a categorization method based on improved sustainability-related qualities and provide general summaries of these tactics. Additionally, we discuss the available empirical evidence for their effectiveness.

\section{Related Work}
\label{sec:related_work}

First, we present some literature regarding the definition and identification of architectural tactics and current developments in academia.

Bachmann et al. \cite{bachmann2003deriving} discuss the relationship between the quality attribute and architectural design while addressing how architectural tactics play an important role in this relationship. Architectural tactics are responsible for adjusting the response of the quality attribute model through changes in architectural design. In other words, architectural tactics are design decisions that influence quality attributes. It is crucial to distinguish tactics from patterns. Here are some key differences:

\begin{enumerate}
    \item A pattern is created using multiple tactics \cite{marquez2023architectural}.
    \item The result of applying a pattern is either benefits or liabilities to quality attributes without going into a detailed explanation, while tactics affect the quality attribute response with a detailed quality attribute model \cite{harrison2010architecture}.
\end{enumerate}

By using architectural tactics, we can meet the requirements of quality attributes \cite{bachmann2003deriving}. The authors present a methodology for depicting architectural tactics in their work. The process can be summarized as follows:

\begin{enumerate}
    \item Find and define a software application scenario, which outlines the source, stimulus, environment, artifact, response, and response measure.
    \item Propose a model that describes how quality attributes are achieved in such a scenario. It contains a reasoning framework, quality attributes, independent parameters, dependent parameters, and response measures.
    \item Select free parameters that can be manipulated among the independent parameters.
    \item Find tactics that affect these parameters.
\end{enumerate}

Osses et al. \cite{osses2018exploratory} focus on architectural patterns and tactics for microservices in scientific literature. They note that while there has been increasing interest in researching microservices in academia following their widespread adoption in industry, there have been few studies on tactics. The authors classify architectural patterns from 44 studies using a specific taxonomy and provide a simple classification of tactics. The study offers insights into the use of taxonomy in previous research and highlights the tendency of academia to combine DevOps, microservices, and IoT in terms of patterns.

Márquez et al. \cite{marquez2019identifying} identified use cases in popular open-source microservice systems and proposed a framework for describing the tactics, with a sole focus on the security aspect. It clearly distinguishes the difference between architectural patterns and architectural tactics. However, it is unclear whether this knowledge can be applied in an industrial setting. The workflow for identifying tactics mentioned in their study could be useful for our work. It provides a systematic approach for extracting tactics from open-source projects and describing them in a framework. This approach combines code interpretation with documentation.

Environmental sustainability is a newly emerged goal in software development, and it encompasses a set of quality aspects, including energy efficiency, resource efficiency, power consumption, greenability, and durability \cite{garcia2018interactions}. While no secondary study exists on the environmental sustainability of microservices, there are several related reviews in the field of cloud applications and other optimizations in microservice systems.

Patel et al. \cite{7155006} presents a study on green computation in the cloud. The review discusses seven applications of green IT, such as environmental sustainability design and eco-labeling, and aligns the objectives of cloud computing with these green IT areas. However, the review only provides a high-level classification of applications and does not delve into specific concerns.

Radu \cite{radu2017green} also conducted a literature review on cloud computing but did not specifically mention microservices. The review discusses research interests related to green cloud computing, focusing on the level of change. It examines whether efforts to improve resource management efficiency occur at the algorithmic or architectural level. The research reveals a shift in the research trend, with a significant improvement in focus on the algorithmic level in recent years, while architectural improvements continue to be overlooked in academia.

Bushong et al. \cite{bushong2021microservice} conducted an overview of microservice challenges. From the performance aspect, they only focus on common performance metrics without mentioning energy efficiency.

Soldani et al. \cite{soldani2018pains} selected literature from various sources, such as tech blogs and videos, which contain a lot of valuable information due to the popularity of microservices in the industry. Although the review covers most stages of developing and maintaining microservices, there is no mention of energy efficiency and sustainability.

Balanza-Martinez et al. \cite{balanzatactics} addressed the approach to improving energy efficiency without specifying a cloud application or microservice. It shows that most methods lack industry interest and involvement.

Panwar et al. \cite{panwar2022systematic} conducted a comprehensive review on improving the energy efficiency of cloud virtual machine systems. The study provides detailed percentages of improvements and notes the measurement methodology.

Pinto et al. \cite{pinto2015refactoring} gather the research on refactoring approaches for optimizing energy consumption at the application level. Although the review does not address microservices, some of its findings, such as API communications, can serve as an inspiring starting point for optimizing microservices in terms of sustainability.

These research studies provide a valuable guide for current software sustainability research. However, they either do not focus on microservices or only select a portion of concepts derived from sustainability. As a result, a literature review to synthesize microservice tactics to improve environmental sustainability would provide more actionable guidance in this field.

\section{Study Design}
\label{sec:design}

\subsection{Research Goal}
\label{sec:research-goal}

As mentioned in Section \ref{sec-intro}, our research aims to synthesize the architectural tactics that improve environmental sustainability in microservices. The growing proportion of emissions from the ICT industry, the increasing popularity of microservices, and foreseeable legislation provide the motivation for our study.

Our research is a rapid review \cite{cartaxo2020rapid} that focuses on providing insightful observations for industry developers within the context of microservice architecture. To introduce the purpose and the general concepts of the rapid review, we will illustrate the differences between a rapid review and a systematic literature review, and note that they are not a replacement for each other. Some key differences are listed below:

\begin{enumerate}
    \item Rapid reviews always focus on a practical problem, while systematic reviews rarely focus on problems extracted from practice.
    \item Rapid reviews provide solutions for industrial practitioners, while systematic reviews mostly provide insights for researchers.
    \item Rapid reviews are in close collaboration with practitioners, while systematic reviews mostly work with researchers.
    \item Rapid reviews are shorter in research time span and more limited in literature search range.
\end{enumerate}

We aim to identify architectural tactics that can enhance environmental sustainability. The industry partner for this rapid review is Software Improvement Group \cite{sig}, which focuses on Green IT and the value that research on microservice sustainability can contribute to the code analysis of existing codebases. Software Improvement Group helps us outline the study proposal towards sustainability tactics and build an experimental environment for future studies based on this rapid review on their cluster.

This leads us to the question of which architectural tactics can be employed in microservices to improve environmental sustainability. Since this review is about the synthesis of tactics, there are no metrics in use.

\subsection{Research Questions}
\label{sec:questions}

The research questions are listed as follows:

\begin{enumerate}
    \item[{RQ}1-] Which architectural tactics to improve the environmental sustainability of microservices can be synthesized from scientific literature?
    \begin{enumerate}
        \item[{RQ}1a-] How can these tactics be categorized?
        \item[{RQ}1b-] What evidence for the effectiveness of these tactics exists?
    \end{enumerate}
\end{enumerate}

\begin{table*}[ht]
  \centering
  \caption{Search Strings and Corresponding Result Counts}
  \begin{tabularx}{\textwidth}{|>{\hsize=1.7\hsize}X|>{\hsize=.3\hsize}X|}
    \hline
    \textbf{Search String} & \textbf{Result Count} \\
    \hline
    allintitle: (microservice OR microservices OR micro-service OR micro-services) energy & 40 \\
    \hline
    allintitle: (microservice OR microservices OR micro-service OR micro-services) green & 7 \\
    \hline
    allintitle: (microservice OR microservices OR micro-service OR micro-services) sustainable & 10 \\
    \hline
    allintitle: (microservice OR microservices OR micro-service OR micro-services) sustainability & 2 \\
    \hline
    allintitle: (microservice OR microservices OR micro-service OR micro-services) carbon & 2 \\
    \hline
    allintitle: (microservice OR microservices OR micro-service OR micro-services) efficiency & 10 \\
    \hline
    allintitle: (microservice OR microservices OR micro-service OR micro-services) efficient & 85 \\
    \hline
    allintitle: (microservice OR microservices OR micro-service OR micro-services) resource & 121 \\
    \hline
    allintitle: (microservice OR microservices OR micro-service OR micro-services) resources & 13 \\
    \hline
    allintitle: (microservice OR microservices OR micro-service OR micro-services) optimization & 87 \\
    \hline
    allintitle: (microservice OR microservices OR micro-service OR micro-services) optimizing & 17 \\
    \hline
    allintitle: (microservice OR microservices OR micro-service OR micro-services) optimize & 6 \\
    \hline
  \end{tabularx}
  \label{tab:search-strings}
\end{table*}

\subsection{Initial search}

The search engine will be Google Scholar \cite{googlescholar}. Although there are multiple sources for finding relevant software engineering studies, such as IEEE Transactions on Software Engineering and Springer, Google Scholar suits our purpose better because it indexes research from most academic journals, conference proceedings, and online repositories, ensuring we obtain a sufficient number of relevant studies on a topic that is not extensively studied. In addition, like other online repositories for academic studies, Google Scholar provides advanced search string patterns such as boolean operators and wildcard matching to make our initial search flexible, and a "cite by" function to facilitate the snowballing process, which we will explore in Section \ref{sec:snowballing}.

Google Scholar distinguishes between search results for singular and plural forms of nouns. So, we search for both "microservice" and "microservices", for example. To avoid an exploding number of irrelevant research, we search for keywords in titles.

Sometimes, people use other terms to indicate microservices, but including them as keywords leads to a high false positive rate. For instance, "container" shows up in microservice research, but it is also jargon in the field of supply chain. "Web application" also shows up in microservice research, but it covers a wider yet different research field. So, we keep only "microservice" with sustainability-related keywords. Of course, we can combine all sustainability-related keywords with "OR" in one Google Scholar search string, but it turns out that Google Scholar does not comprehend long search strings very well. As a result, we distribute different sustainability-related keywords into a new query and see if the result is too large to handle. These search strings are iteratively refined based on checking the true positives and relevance of the search results.

The search strings and their corresponding result sizes are shown in Table \ref{tab:search-strings}. Fortunately, all result sizes can fit into our subsequent study.

\subsection{Application of selection criteria}
\label{sec:application of selection criteria}

We have inclusion criteria and exclusion criteria to eliminate unrelated research in Google Scholar query results. The purpose is to reduce the number of studies and find research that can help us synthesize the target tactics more effectively. When we refer to target tactics, we are talking about architectural tactics that enhance the quality attributes of environmental sustainability through design decisions. Therefore, we introduce the following inclusion criterion:

\begin{enumerate}
\item[{I}1-] Describes at least one design decision to improve the environmental sustainability of microservices.
\end{enumerate}

The exclusion criteria involve language restrictions. Additionally, we solely concentrate on peer-reviewed publications, disregarding gray literature, as its quality and effectiveness are not evaluated, and the verification process is beyond the scope of our research.

Since the number of research studies extracted from the Google Scholar query result is already limited, we only focus on primary research. Secondary research studies often overlap significantly with each other and primary research. The exclusion criteria are listed as follows:

\begin{enumerate}

\item[{E}1-] Written in a language other than English.

\item[{E}2-] Not a peer-reviewed scientific publication, e.g., gray literature.
  
\item[{E}3-] Secondary studies.

\item[{E}4-] Not accessible by us.
  
\end{enumerate}

\subsection{Snowballing}
\label{sec:snowballing}

After obtaining the results by applying inclusion and exclusion rules, which is called the start set of snowballing \cite{wohlin2014guidelines}, we perform one round of forward snowballing and backward snowballing. This helps us identify neglected research that was not found through our initial keyword search. The backward snowballing relies on the references mentioned in the start set. The forward snowballing aims to find research that references the paper in the start set. We use Google Scholar's "cite by" function to implement forward snowballing.

\begin{figure}[ht]
\centering
\includegraphics[width=0.4\textwidth]{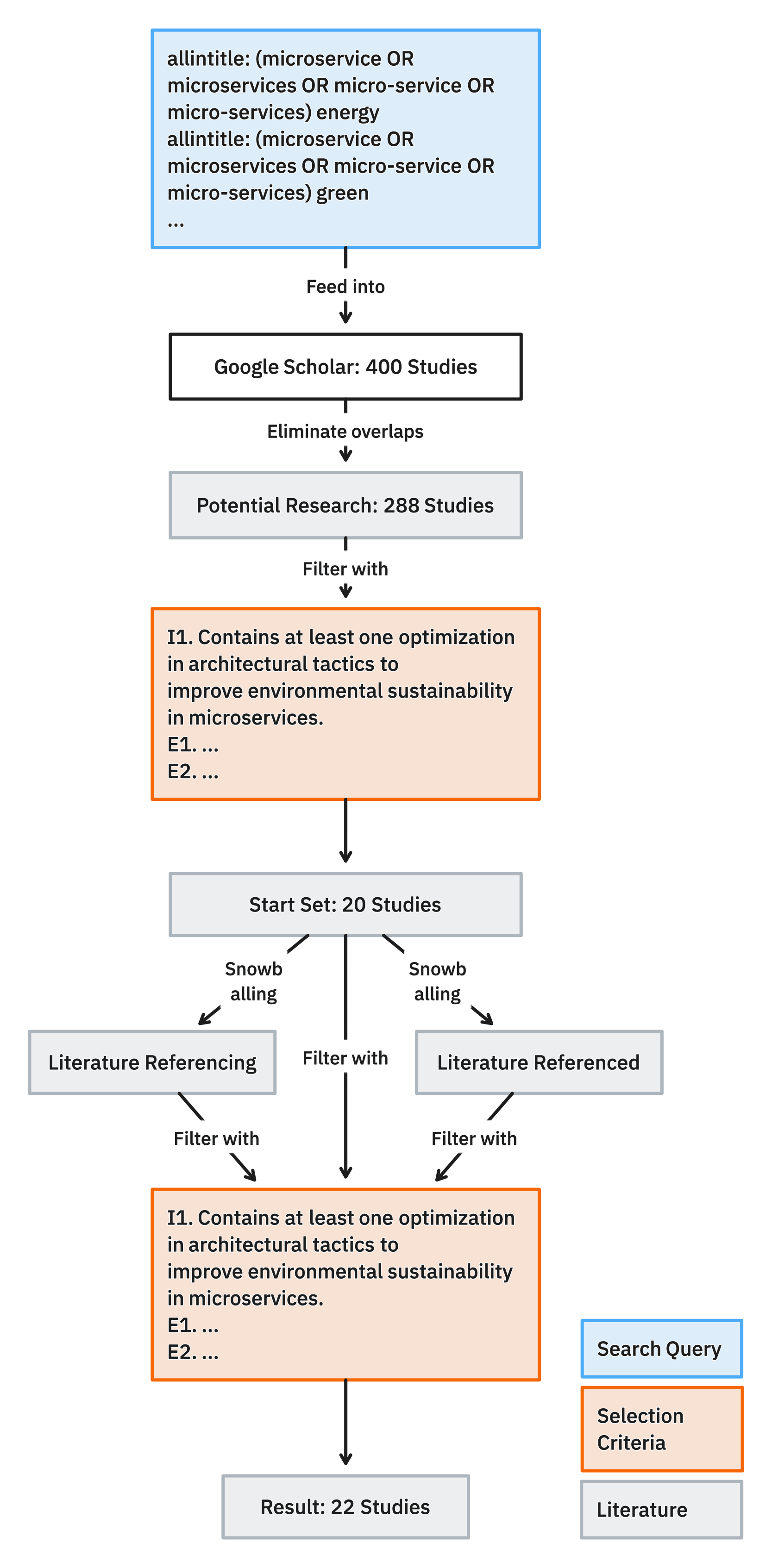}
\caption{The process of gathering relevant literature}
\label{fig:selection flowchart}
\end{figure}

One round of snowballing ensures that the result set remains manageable. Subsequently, we reapply the inclusion and exclusion rules to maintain the relevance of the results to our study. There are 22 studies that fulfill the condition at the end. The complete process of literature selection is shown in Figure \ref{fig:selection flowchart}.

\subsection{Data Extraction}
\label{sec:data extraction}

In this section, we discuss the data needed to be collected from the studies, including context, intervention, study methods, and outcome of the research \cite{lisections}. They are needed in the process of data synthesis mentioned in Section \ref{sec:data synthesis}.

\begin{itemize}
    \item \textbf{Context}: Most of the studies collected focus on different application contexts. For example, some propose new methods to reduce GPU cluster energy consumption, while there is also a significant amount of research on Internet of Things microservices. The application domain will be extracted from the research. Both practitioners and we need a context to determine the application range of the target study so that our synthesized tactics can be properly adopted in the future.
    \item \textbf{Intervention}: To conduct an optimization in a microservice system, an intervention is implemented. It includes the affected part in the microservice system, how the intervention is implemented, and how the intervention affects the outcome. We use the term "affected part" because it is not limited to a specific level in the microservice system. It can be the API gateway or the whole network topology of how microservices are connected. This type of data extraction has already been done partly because we need to identify studies that fulfill inclusion criterion I1 mentioned in Section \ref{sec:application of selection criteria}.
    \item \textbf{Study Methods and Outcome}: There are different approaches to prove the effectiveness of tactics. We will first extract the evidence-proofing process from the studies, which shall include with which tools the evidence-proofing experiment is set up and conducted, and how the conclusion is drawn. We address the tools and the setup of the experiment because it is beneficial for Software Improvement Group's interest in automating the testing environment for its microservices, and it facilitates our future research on microservices in collaboration with the Software Improvement Group. The evidence study provides us with insight into the preference of academia on microservice-related experiments and what kinds of results the current toolchain of experiments can achieve.
\end{itemize}

\subsection{Data Synthesis}
\label{sec:data synthesis}

This section provides an overview of the process of how we synthesize information relevant to research questions from the data we collect from the studies. 

First, we will synthesize what the tactics are about in each piece of literature we found to answer research question RQ1. Then, we will codify the architectural tactics into the different sustainability aspects, whose categorization is shown in Section \ref{sec: Codifying Architectural Tactics}. The application context of the microservices mentioned in the research will also be noted. These constitute the answer to research question RQ1a. Finally, to facilitate future studies on the architectural tactics on microservices, we synthesize study methods and outcomes to answer RQ1b.

\subsubsection{Synthesizing Architectural Tactics (RQ1)}
\label{sec:Synthesizing Architectural Tactics (RQ1)}

Since we have extracted the intervention from the studies, we can now identify the context, the goal, design decisions, the target quality attribute, the quality attribute model, and the quality attribute response in the studies. We combine the information and synthesize architectural tactics used in the research to form the description of tactics \cite{bachmann2003deriving}.

\subsubsection{Categorizing Architectural Tactics (RQ1a)}
\label{sec: Codifying Architectural Tactics}

Since the goal of architectural tactics is to satisfy quality attributes, we need to log the corresponding quality aspects. In our case, the goal of applying architectural tactics is to improve environmental sustainability, and the quality aspects we want to satisfy are relevant to environmental sustainability. To be more specific, sustainability aspects will be classified into the following categories \cite{garcia2018interactions} so that they can be linked with architectural tactics found in the research. To assist in future studies within Software Improvement Group, we also introduce the concepts of sustainability and carbon efficiency. Energy efficiency and power consumption are two terms describing environmental sustainability sometimes used interchangeably, so we merge them into the term "energy efficiency."

\begin{itemize}
    \item \textbf{Energy Efficiency}: Comparison of the effective output to the energy input \cite{gregor2015briefing} or the energy used within a timeframe \cite{harris2015digital}.
    \item \textbf{Resource Efficiency}: Comparison of the effective output to the resource utilization. The resource refers to hardware components such as CPU, memory, and GPU.
    \item \textbf{Greenability}: The ability to optimize the energy and resources over a longer period of time \cite{martinez2023towards}.
    \item \textbf{Carbon Efficiency}: Comparison of the effective workload to the carbon dioxide emission \cite{garg2011green}.
\end{itemize}

To complete the description of architectural tactics, we also need the implemented design decisions and quality attribute responses. Naturally, the quality attribute response can be related to our research question 1b. It is usually evaluated by changing one input parameter in the quality attribute model and measuring the variation of output parameters.

Additionally, as we have mentioned context information in Section \ref{sec:Synthesizing Architectural Tactics (RQ1)}, it is also an important categorization, which is further discussed in Section \ref{sec:Categorization Based on Context} based on the researches we found because we have no predefined categorization like sustainability aspects.

\subsubsection{Codifying Evidence for the Effectiveness (RQ1b)}

Since we have extracted study methods and outcomes from the studies, we are able to show with which tools and infrastructure the studies are conducted and how conclusions are drawn. For example, we can use them to find out the commonly used monitoring stack in the microservice system and how it records the measurement of sustainability aspects. Apart from that, the statistical analysis of effectiveness is important for later studies on proving the effectiveness of architectural tactics. It indicates how the monitoring system should be set up and what data we should gather in a new architectural tactics-related study. We therefore classify statistical methods of the studies into descriptive statistics, graphical representations, and hypothesis testing.

\subsection{Threats to Validity}

We use the categorization proposed by Zhou et al. \cite{7890583} to present the threats to validity in our rapid review. Validity can be broken down into construct validity, internal validity, external validity, and conclusion validity. Construct validity refers to applying correct operational measures for the target concept. Internal validity refers to the correct form of the relationship between one condition and another. External validity refers to the correct generalization of the research domain. Conclusion validity refers to the repeatability of the conclusion. In the following subsections, we will briefly present all threats to validity and classify them according to the validity they might breach. We use the aspect of threats proposed by Zhou et al. \cite{7890583} as subsection titles to better present the threats to validity in our review.

\subsubsection{Incomplete and Incorrect Search for Studies}

Although we have tried the selection criteria with different keywords and search queries, there is still a high number of false positive search results. Many results are related to the Internet of Things and microservice management systems, which are not relevant to our study. It is possible that some literature is also omitted in this process. This poses threats to construct validity and internal validity.

\subsubsection{Inadequate size and number of samples}

Although we have tried different search strings and selection criteria, there is an inadequate number of studies in the related field. With the sample size being so small, they may not accurately reflect the trends and research hotspots in academia. This impacts the accuracy of results and poses a threat to internal validity.

\subsubsection{Misclassification of primary studies and lack of standard languages and terminologies}

Due to the fact that the categorization mentioned in our rapid review is standardized, the differences in the definition of sustainability-related terms may affect the construct validity and internal validity. For example, we use definitions of sustainability aspects synthesized from the literature review done by García-Mireles et al \cite{garcia2018interactions}. The definition of architectural tactics, which is the main starting point of the review, may appear to be defined differently in other studies. The meanings of carbon efficiency and energy efficiency are different in different studies, especially in research whose main topic is not environmental sustainability, which makes the results of these studies incomparable. This ultimately threatens the conclusion validity as well.

\subsubsection{Unsatisfactory data synthesis}

We primarily follow the style of describing architectural tactics proposed by Bachmann et al. \cite{bachmann2003deriving}. While the description is concise and contains key aspects of architectural tactics, such as context, target quality attribute, the parameters of the quality attribute, and the purpose of this tactic, it lacks some information compared to the definition of architectural tactics. It poses threats to internal validity.

\subsubsection{Subjective interpretation of the extracted data}

Many data extracted need the involvement of subjective interpretation, such as intervention and outcome. This affects the synthesis of tactics and the evidence of effectiveness. It poses threats to internal validity and conclusion validity.

\section{Results}

\subsection{RQ1: Architectural Tactics Synthesized from Literature}

We have summarized the architectural tactics gathered from the research in Table \ref{tab:arch-tactics}. The tactic details used in each research are different, but some of them can be generalized into several large categories. These tactics can be synthesized in general as follows:

\begin{enumerate}

\item[{T}1-] \textbf{Use elastic containers}: Traditional containers are allocated a fixed amount of resources before deployment, which makes it impossible to utilize idle resources and makes resource estimation for each container before deployment a complicated and challenging task. Elastic containers allow dynamic allocation of resources based on monitoring each container's resource utilization.

\item[{T}2-] \textbf{Distribute pods to different nodes to take advantage of node properties}: Nodes have their own properties, such as the energy grid they are powered by (solar energy, thermal power, etc.), or the acceleration for specific software that some pods are based on. For example, GPU acceleration servers are good at hosting machine learning pods. This makes a difference when assigning different pods to various nodes. Sometimes, it is simply a matter of load balancing to distribute pods across different nodes.
  
\item[{T}3-] \textbf{Distribute microservices to different containers based on affinity}: Sometimes it is beneficial to group similar microservices in the same container, as they share similar underlying software dependencies or functions. This approach also allows for aggregating API requests that access similar resources to the same container.

\item[{T}4-] \textbf{Use predictive autoscaling algorithm}: Autoscaling involves adjusting resources allocated to microservices in response to changes in demand. Typically, this entails adjusting the number of pod replicas or hardware resources assigned to them. Some advanced autoscaling algorithms incorporate prediction algorithms to optimize resource allocation proactively. Others consider new factors, such as carbon emissions, which serve as the basis for these prediction algorithms. Currently, the autoscaling algorithm is usually proposed based on machine learning and deep learning techniques, or concepts borrowed from control theory. Sometimes, in a more fine-grained control approach, developers have designed different versions of containers that behave differently in response to varying loads of demands. The autoscaling algorithm is responsible for switching between them.

\item[{T}5-] \textbf{Use energy-efficient hardware}: This tactic is based on adapting to utilize newly proposed hardware to deploy microservices. For example, a network interface card with a microprocessor may not be able to deploy large cloud applications, but it is possible for them to deploy microservices in some cases \cite{liu2019e3} and reduce latency when executing API calls. The purpose of this tactic is to reduce energy consumption, especially when it comes to microservice-specific features like sending and receiving API requests frequently.

\item[{T}6-] \textbf{Use different proxies for different demands}: Since microservices heavily rely on API gateways for communication \cite{nathaniel2023istio}, proxy selection is crucial. Some proxies are efficient for handling a large number of requests, while others are power-saving when idle.
  
\end{enumerate}

We will refer to tactics type as T\{1-6\} in the following sections.

\begin{table*}[ht]
  \centering
  \caption{Architectural Tactics Collection}
  \begin{tabularx}{\textwidth}{|>{\hsize=.65\hsize}X|>{\hsize=.45\hsize}X|>{\hsize=.5\hsize}X|>{\hsize=.4\hsize}X|}
    \hline
    \textbf{Tactics} & \textbf{Sustainability Aspects} & \textbf{Research} & \textbf{Context} \\
    \hline
    T1: Use elastic containers & Resource efficiency
    & \cite{cusack2019efficient} & General \\
    \hline
    T2: Distribute pods to different nodes to take advantage of node properties & Energy efficiency, Carbon efficiency
    & \cite{de2021revisiting}, \cite{10.1145/3412841.3441888}, \cite{valera2022pisco}, \cite{saboor2022enabling}, \cite{dell2022adaptive} & General, Decentralized network \\
    \hline
    T3: Distribute microservices to different containers based on affinity & Resource efficiency
    & \cite{10305917} & General \\
    \hline
    T4: Use predictive autoscaling algorithm & Energy efficiency, Carbon efficiency, Resource efficiency
    & \cite{gebreweld2023evaluating}, \cite{saboor2022enabling}, \cite{dell2022adaptive}, \cite{fu2021adaptive}, \cite{hossen2022practical}, \cite{harr2019efficient}, \cite{baarzi2021showar}, \cite{li2022score}, \cite{santos2023gym}, \cite{zhang2021sinan}, \cite{al2023proactive}, \cite{luo2022erms}, \cite{li2020amoeba}, \cite{jagtap2020optimal} & General, Serverless platform \\
    \hline
    T5: Use energy-efficient hardware & Energy efficiency
    & \cite{liu2019e3}, \cite{khairy2022simr} & Hardware \\
    \hline
    T6: Use different proxies for different demands & Resource efficiency 
    &  \cite{nathaniel2023istio} & General \\
    \hline
  \end{tabularx}
  \label{tab:arch-tactics}
\end{table*}

\subsection{RQ1a: Architectural Tactics Categorization}

\subsubsection{Categorization Based on Sustainability Aspects}

In our research, we identified sustainability aspects as the optimization goal, as mentioned in Section \ref{sec: Codifying Architectural Tactics}. These aspects include energy efficiency, resource efficiency, greenability, and carbon efficiency. However, our collected literature only mentions energy efficiency, resource efficiency, and carbon efficiency. There is currently no discussion about greenability, as this concept has not been widely accepted in academia.

Energy efficiency refers to the ratio of the amount of work done to the energy consumption. It is often regarded as energy consumption, power consumption, or power efficiency in studies. Thus, there are actually some differences regarding the definition of energy efficiency among studies. Many studies in this category always start with an energy efficiency model, which acts as a complex result representing the energy efficiency built from basic measurements. However, many studies do not provide detailed information on how these measurements are obtained.

Resource efficiency refers to the utilization of memory and CPU. Some research also addresses concerns about network latency. Researches that require the accuracy of experiments also consider CPU frequency. There is not much difference in the definition of resource efficiency among the selected researches.

Carbon efficiency generally refers to the ratio of the amount of work done to the emission of carbon dioxide. The definition of the amount of work varies from research to research. The emission of carbon dioxide is sometimes substituted with the emission of greenhouse gases.

Among the three sustainability aspects mentioned, resource efficiency is the most researched, accounting for 63.64\% of all studies. Energy efficiency-related research makes up 33.33\% of the total, while carbon efficiency research accounts for 12.5\%.

\begin{figure}[ht]
\centering
\includegraphics[width=0.45\textwidth]{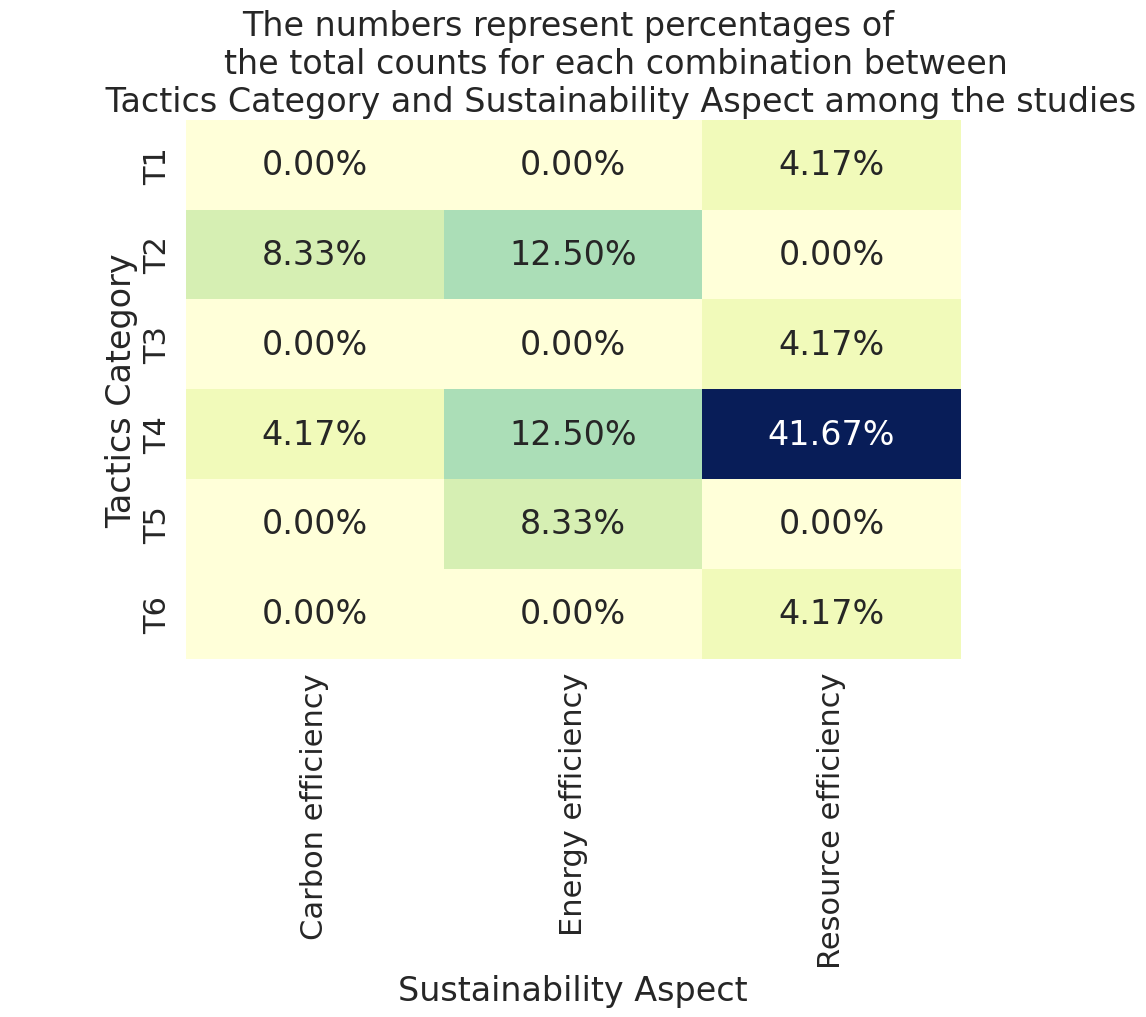}
\caption{Tactics and Sustainability Aspects Research Heatmap}
\label{fig:tactics-sustainability-heatmap}
\end{figure}

Figure \ref{fig:tactics-sustainability-heatmap} shows the percentage of each combination of tactics used and target sustainability aspect in research. It is evident that employing predictive autoscaling algorithms in the realm of resource efficiency is a prominent area of research. This is possibly due to advancements in machine learning and deep learning, which offer the potential to enhance existing scaling and scheduling algorithms, as indicated by these studies. There is also some research regarding the distribution of pods on different nodes when energy efficiency is the optimization goal. Other combinations of tactics and researched sustainability aspects are not studied extensively.

\subsubsection{Categorization Based on Context}
\label{sec:Categorization Based on Context}

There are different contexts that these research studies address. Most research does not specify the limits of its application context, which we label as \textit{General} in Table \ref{tab:arch-tactics}.

Microservices can be deployed in a decentralized network, which has advantages over a centralized network, such as scalability and privacy, and one research is specifically about that. It is labeled as \textit{Decentralized network} in Table \ref{tab:arch-tactics}. What makes it different from the general context is that the decision on which node in a decentralized network to execute the microservice matters. It is related to the requirement of microservice itself and the configuration of nodes and connectivity of nodes in a decentralized network.

Some architectural tactics focus on the improvement of hardware algorithms, and it's worth categorizing because it usually requires data center construction with specific hardware, and the optimizations that the tactics introduce are hardware-dependent. It is labeled as \textit{Hardware} in Table \ref{tab:arch-tactics}. For instance, Khairy et al. \cite{khairy2022simr} proposed a new requests scheduling algorithm but it only be implemented on top of the request processing unit proposed in the study. Hardwares like request processing unit make the optimization of microservice because they take advantage of features that are prominent on microservices. The request processing unit in this case identifies similar and frequent requests between microservices.

Some designs aim to be deployed on cloud edge, which is increasingly popular in the IoT industry, such as smart vehicles. Special features like low computational resources on the edge and quick responses by edge devices pose new challenges to microservice design, thus, we labeled them as \textit{Edge computing}.

There are also discussions about the serverless platform, which is the new standard of cloud providers, we labeled the tactics aiming to work on serverless platforms as \textit{Serverless platform}. Serverless platform focuses on low-load microservices, and may not meet users' need as the requests for microservices grow larger. It poses new challenges to microservice design under low resource restriction.

Among the research we gathered, 77.27\% of studies do not specify the exact application context. 4.55\% of studies apply only to decentralized networks, edge computing, and serverless platforms respectively. 9.09\% of studies are about hardware innovation.

\subsection{RQ1b: Evidence for the Effectiveness of Tactics}
\label{sec:evidence}

Most studies include statistical results to demonstrate the effectiveness of implementing sustainable tactics.

In the initial three subsections, we examine evidence supporting the effectiveness of tactics across various sustainability aspects. Due to commonalities in experiment setups and measurement units among each category, we discuss a general experimental approach derived from selected studies. We present measurement units which are valuable for future analysis of microservice sustainability at Software Improvement Group and provide guidance on selecting measurement tools for future research.

In Section \ref{sec:Statistical Methods}, we summarize the statistical methods used in studies. In Section \ref{sec:Experiment Setup}, we outline a common experimental toolchain and address the deficiencies in the measurements. These sections are valuable for report writing and experiment setup, representing a widely accepted paradigm in this field.

\subsubsection{Research Related to Energy Efficiency}

\begin{table}[ht]
  \centering
  \caption{Tactics for Improving Energy Efficiency, Related Measurement Units and Statistical Methods}
  \begin{tabularx}{0.48\textwidth}{|>{\hsize=.12\hsize}X|>{\hsize=.44\hsize}X|>{\hsize=.44\hsize}X|}
    \hline
    \textbf{Tactics} & \textbf{Measurement Units} & \textbf{Statistical Methods} \\
    \hline
    T2 & \textit{kWh}, \textit{Watt}, \textit{Joule}
    & Graphical representation, Descriptive statistics \\
    \hline
    T4 & \textit{Joule}
    & Hypothesis testing \\
    \hline
    T5 & \textit{kWh}
    & Graphical representation, Descriptive statistics \\
    \hline
  \end{tabularx}
  \label{tab:Tactics Improving the Energy Efficiency and Related Measured Unit and Statistical Methods}
\end{table}

In the evaluation of energy efficiency, most studies compare an optimized version of a microservice system with a naively implemented one. For example, Gebreweld \cite{gebreweld2023evaluating} compares the carbon-aware autoscaler to the horizontal pod autoscaler using a t-test. The study concludes that the new autoscaler significantly saves more power than the traditional horizontal pod autoscaler. Most studies' evidence is based on practical tests, while a few of them rely on theoretical computation and simulation. For instance, Valera et al. \cite{valera2022pisco} compares an optimized scheduler with a basic one but does not directly state the comparison result or provide statistical data, as the focus of the research is on the usage of the deployment simulator proposed in the study.

Researchers use units like \textit{kWh}, \textit{Watt}, and \textit{Joule} to show the energy efficiency of a microsystem. Tactics for improving energy efficiency, related measured units, and statistical methods are shown in Table \ref{tab:Tactics Improving the Energy Efficiency and Related Measured Unit and Statistical Methods}.

Sometimes, to make graphical representations and descriptive statistics more prominent in terms of the comparison between the naive version and optimized version of microservices, researchers choose to use stress tests. For example, Valera et al. \cite{valera2021energy} discusses innovative methods to determine hardware energy efficiency based on hardware features such as running frequency. The experiment includes scalability and stress tests, both demonstrating a decrease in power consumption.

In terms of repeatability and general representation of the experiment, some choose to use an open-source toolchain. Dell et al. \cite{dell2022adaptive} have a comprehensive and reproducible experimental environment. The study mentions the platform and open-source tools like Locust and Prometheus used for creating test cases and monitoring consumption, showing real-time improvements in energy efficiency. We will explore more on this topic in Section \ref{sec:Experiment Setup}.

\subsubsection{Research Related to Resource Efficiency}

\begin{table}[ht]
  \centering
  \caption{Tactics for Improving Resource Efficiency, Related Measurement Units and Statistical Methods}
  \begin{tabularx}{0.48\textwidth}{|>{\hsize=.12\hsize}X|>{\hsize=.44\hsize}X|>{\hsize=.44\hsize}X|}
    \hline
    \textbf{Tactics} & \textbf{Measurement Units} & \textbf{Statistical Methods} \\
    \hline
    T1 & \textit{Percentage (CPU Utilization or Memory Utilization)}
    & Descriptive statistics \\
    \hline
    T3 & \textit{Gb (or Mb)}, \textit{Cores (CPU)}
    & Graphical representation, Descriptive statistics \\
    \hline
    T4 & \textit{Percentage (CPU Utilization or Memory Utilization)}, \textit{Second (CPU throttling time)}
    & Graphical representation, Descriptive statistics, Hypothesis testing \\
    \hline
    T6 & \textit{Percentage (CPU Utilization or Memory Utilization)}
    & Graphical representation, Descriptive statistics \\
    \hline
  \end{tabularx}
  \label{tab:Tactics Improving the Resource Efficiency and Related Measured Unit and Statistical Methods}
\end{table}

In terms of resource efficiency evaluation, most studies also compare a proposed new method with a selected baseline using statistical analysis. Resource utilization is easier to measure than carbon efficiency and energy efficiency because the measured unit, CPU utilization, and memory utilization, can be directly obtained from the Linux APIs. These values can be extracted by multiple monitoring tools and presented in intuitive graphical representations.

The most commonly used measurements related to resource efficiency are CPU utilization and memory utilization. More accurate units are employed when tasks are related to maintaining a specific quality of service in a study. The tactics for improving resource efficiency and the related measured units and statistical methods are shown in Table \ref{tab:Tactics Improving the Resource Efficiency and Related Measured Unit and Statistical Methods}.

Generally, there are two ways to compare the naive version and optimized version of microservices. First, it can be done by increasing the test load and observing how the two microservice systems react to the load change while keeping the hardware resources unchanged. Second, allow the hardware configuration to change freely while adjusting the test load. The second setup is more akin to an elastic server and may reflect the hardware scheduling regarding how the microservice system adapts to the requirements when the load is either extremely small or extremely large.

\subsubsection{Research Related to Carbon Efficiency}

\begin{table}[ht]
  \centering
  \caption{Tactics for Improving Carbon Efficiency, Related Measurement Units and Statistical Methods}
  \begin{tabularx}{0.48\textwidth}{|>{\hsize=.12\hsize}X|>{\hsize=.44\hsize}X|>{\hsize=.44\hsize}X|}
    \hline
    \textbf{Tactics} & \textbf{Measurement Units} & \textbf{Statistical Methods} \\
    \hline
    T2 & \textit{Gram (or Ton, Kilogram)}
    & Graphical representation, Descriptive statistics \\
    \hline
    T4 & \textit{Gram (or Ton, Kilogram)}
    & Hypothesis testing \\
    \hline
  \end{tabularx}
  \label{tab:Tactics Improving the Carbon Efficiency and Related Measured Unit and Statistical Methods}
\end{table}

As for carbon efficiency, it is significantly less discussed than the other two aspects. Relevant studies all use weight units as the measurement of carbon efficiency, but their measurement models for carbon efficiency differ. For example, Saboor et al. \cite{saboor2022enabling} use calculated carbon efficiency based on the assumed locations of data centers, while Gebreweld \cite{gebreweld2023evaluating} uses WattTime \cite{watttime}, a electricity grid query API service, to fetch carbon emission data from the electricity grid. In short, none of them measure the carbon emissions directly, and they also mention the difficulty in doing so. These works all try to prove their approach after applying new tactics to improve carbon efficiency by comparing the calculated carbon efficiency data.

Tactics for improving the carbon efficiency and related measurement units and statistical methods are shown in the table \ref{tab:Tactics Improving the Carbon Efficiency and Related Measured Unit and Statistical Methods}.

\subsubsection{Statistical Methods}
\label{sec:Statistical Methods}

Among the energy efficiency optimization-related research, 12.5\% of studies show improvement with hypothesis testing, 25\% of studies show improvement with graphical representation, and 62.5\% of studies show improvement with descriptive statistics. In terms of carbon efficiency-related research, 100\% of the studies demonstrate improvement through graphical representation. Among the research studies aimed at improving resource efficiency, 57.14\% of these studies use descriptive statistics to demonstrate their effectiveness. 42.86\% of these studies employ graphical representation to substantiate their claims.

We can observe that the statistical methods used in the experiments are largely affected by the toolchain researchers can use. For example, when a monitoring system that can export energy efficiency out of the box is used, researchers consistently demonstrate the changes in energy efficiency with graphical representations in the study. The phenomenon will be further discussed in the next section.

\begin{figure}[ht]
\centering
\includegraphics[width=0.49\textwidth]{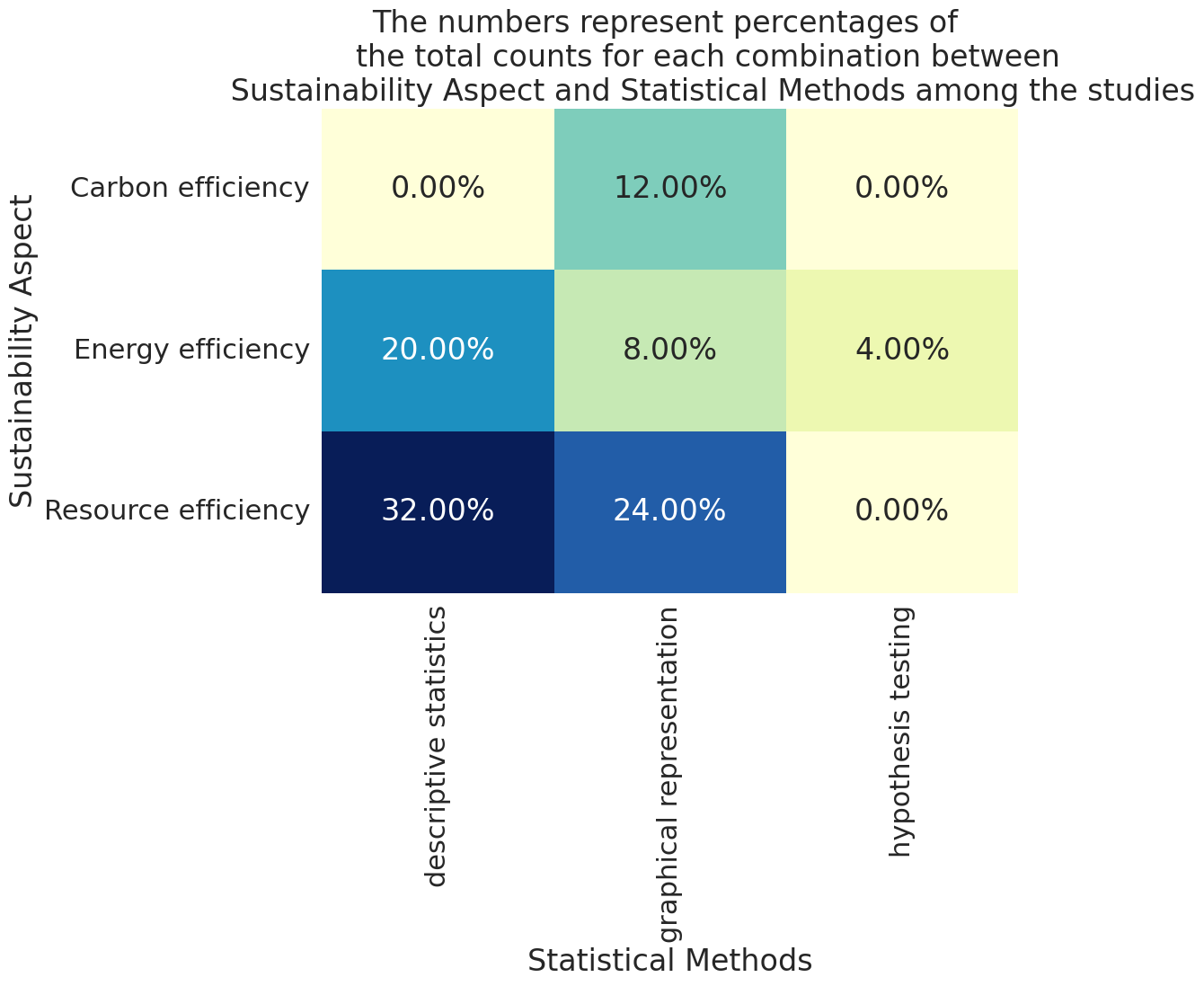}
\caption{Sustainability Aspect and Statistical Methods Research Heatmap}
\label{fig:QA and Evidence Research Heatmap}
\end{figure}

\subsubsection{Experiment Toolchain}
\label{sec:Experiment Setup}

Since the time spans of many experiments are very long, and the extraction of data from microservices is not as convenient as with a local experiment, researchers always seek to generate the statistical representation in an automatic way.

Most experiments were conducted on large cloud provider platforms, namely Google Cloud \cite{googlecloud} and Microsoft Azure \cite{azure}. Some researchers perform experiments on local clusters, which may affect the repeatability of the evidence and conclusions. Docker \cite{docker} and Kubernetes \cite{kubernetes} are becoming the standard for experimenting with microservices. Along with them, Grafana \cite{grafana} and Prometheus \cite{prometheus}, popular monitoring tools in the industry, play an important role in collecting data and plotting graphs of key metrics.

Because of the ease of measurement and the support of hardware utilization APIs, using descriptive statistics and graphical representations to demonstrate effectiveness in resource efficiency is one of the most commonly used statistical methods. With the assistance of popular cloud monitoring tools, exporting the energy efficiency graphical representation is gradually becoming popular in research, as shown in Figure \ref{fig:QA and Evidence Research Heatmap}. Most energy efficiency evaluations are exported by Kepler \cite{amaral2023kepler}, which serves as a power consumption estimator based on trained models. Other methods are also based on an estimation of energy efficiency from factors like the time a microservice is running and the number of nodes in use with different formulas. Carbon efficiency is mostly based on the geolocation of nodes and queries the theoretical carbon emission of the local grid. In conclusion, carbon efficiency and energy efficiency are not direct measurements in these studies, and direct measurement is challenging. There may be a gap between the actual carbon efficiency and energy efficiency and the calculated ones, but there is no research on studying the difference or how to fill the gap between theory and reality.

\section{Discussion} 

In this section, we provide the interpretation of the results' meaning and the implications for practice and research.

\textbf{The ambiguity of the definition for carbon efficiency and energy efficiency models and the difficulty in measurement hinder researchers from delving deep into the relevant tactics.} An indispensable part of architectural tactics is deciding on a model that describes how quality attributes are achieved, but there is no widely agreed-upon model for evaluating the carbon efficiency and energy efficiency of microservice systems. Furthermore, the majority of related experiment statistics are dependent on indirect, unverifiable estimations. Researchers need to collaborate with multiple organizations to obtain accurate data. For instance, the issue becomes exceptionally complex when dealing with carbon efficiency measurements because carbon emissions can be divided into direct emissions by the microservice owner and indirect emissions by electricity producers or other aspects \cite{gebreweld2023evaluating}. As such, local grid data APIs might not be sufficient to achieve industry-level optimization. Future research should prioritize verifying the accuracy of these estimations within the proposed models and thereby improve the robustness of studies.

\textbf{Interdisciplinary borrowings offer a pathway for innovative research on architectural tactics.} Many fast-growing research fields, like deep learning, contribute to developing new quality attribute models and scaling algorithms. Simultaneously, mature fields like environmental sustainability and control theory offer valuable principles for constructing quality attribute models and fundamental definitions for new research in microservice sustainability. Conducting literature reviews on environmental sustainability research in future studies can help address the significant discrepancies in measurements and estimations by proposing new measurement models \cite{GULDNER2024402} of energy efficiency and carbon efficiency, and discovering new architectural tactics to improve microservice sustainability.

\textbf{Many research studies predominantly focus on exploring application-level scenarios rather than delving into architectural innovation in the field of environmental sustainability.} Within these studies, various autoscaling and scheduling algorithms have been proposed, each grounded in different theoretical frameworks such as statistical methods, deep learning, or novel factors like carbon emissions. Despite the diverse theoretical underpinnings, these algorithms collectively operate within a comparable architectural framework. There are not enough new proposals regarding key aspects of architectural tactics like quality attribute models and design decisions.

\section{Conclusion}

In this rapid review, we synthesized information from 22 studies about architectural tactics for enhancing the environmental sustainability of microservice systems. These tactics are cataloged based on sustainability aspects (e.g., carbon efficiency, energy efficiency, resource efficiency) they aim to optimize and their application contexts. We applied literature selection techniques such as keyword search and snowballing to discover the relevant sources.

Our findings reveal six main architectural tactics. Among different sustainability aspects, resource efficiency emerged as the most prevalent because of its measurability. Each of the included studies provides evidence of the effectiveness of these optimization tactics. Statistical methodologies in these studies include graphical representations, descriptive statistics, and hypothesis tests, presenting the advantageous outcomes of applying these architectural tactics.

Findings reveal that the predictive autoscaler is the most commonly employed tactic, largely due to advancements in machine learning and deep learning fields. Containerization tools such as Docker and Kubernetes, along with monitoring stacks like Grafana and Prometheus, were prevalent, establishing a common infrastructure for this kind of research.

There are several potential directions for future research. First, we noticed a research gap in integrating greenability sustainability aspects into architectural tactics. Also, we observed insufficient research on energy efficiency and carbon efficiency in microservices, primarily due to the lack of measuring methodologies and universally-accepted definitions. Thus, future research in these areas could make considerable contributions to the field. Further, borrowing knowledge from other areas, such as environment and control theory, could enrich architectural tactics.

The architectural tactics that improve microservice sustainability remain mostly unexplored territory. While we provide a review of current knowledge, it also points to the need for efforts from both academia and industry to improve the environmental sustainability of microservices. An online evidence briefing is available for industrial practitioners \cite{xiao_2024_10804901}, providing an at-a-glance view of this rapid review.

\bibliographystyle{ACM-Reference-Format}
\bibliography{bibliography}

%

\end{document}